\documentclass[aps,showpacs,twocolumn]{revtex4}

\usepackage{graphics}
\input epsf

\begin{document}

\title{Towards fabrication of ordered gallium nanostructures by laser manipulation of neutral atoms: study of self-assembling phenomena}
\author{B. Fazio$^{1}$, O.M. Marag\`o$^{1}$, E.
Arimondo$^2$, C. Spinella$^3$, C. Bongiorno$^3$, and G.
D'Arrigo$^3$}

\address{(1) CNR-Istituto per i Processi Chimico-Fisici sez.
Messina, via La Farina 237, I-98123 Messina, Italy.\\(2) INFM,
Dipartimento di Fisica, Universit\`a di Pisa, via Buonarroti 2,
I-56127 Pisa, Italy.\\
(3) CNR-Istituto per la Microelettronica e microsistemi (IMM- sez.
Catania), stradale Primosole, 50, 95121 Catania, Italy.}

\date{\today}

\begin{abstract}
Surface diffusion has an impact on the lateral resolution of
nanostructures in bottom-up atom nanofabrication.  In this paper
we study the effects of the gallium atoms self-assembled on
silicon surfaces (100) patterned with trenches at different
slopes. These particular substrate morphologies have been made to
enable an effective deposition rate variation along the surface.
In this way we experimentally mimic the effect of the atomic flux
modulation created by standing wave during an atom nanofabrication
experiment. Even if we observe self organization of gallium atoms
on the surface, we conclude that  the nano-islands are not
affected by surface diffusion processes and the effective
variation of the deposition rate per unit area is the dominant
factor affecting the growth differences along the surface. This
result demonstrates that the gallium atoms self-organization
should not prevent the observation of  a periodic nano-patterning
created by atom nano-fabrication techniques.
\end{abstract}

\maketitle

\section{Introduction}
In recent years a progress towards the fabrication of
nanostructures was made through the laser cooling and manipulation
of neutral atomic beams indicated as Atom Nano-Fabrication
(ANF)~\cite{Meschede}. The ultimate resolution limit in ANF is
associated with the quantum-mechanical wave-like nature of atoms.
It has been shown that highly collimated beams of laser cooled
atoms can be realized with a typical de Broglie wavelength much
smaller than the spacing between atoms in a solid. Thus ordered
nanostructures can be grown at the single-atom scale. The
technique is based on two steps, first the use of laser cooling
methods for the high collimation of the atomic beam, second the
focusing of atoms through a laser beam in standing wave
configuration (light mask). This process produces an ordered
pattern with a precise spacing determined by the laser wavelength.
Arrays of lines and dots have been already produced using few
different atomic species. An important breakthrough for industrial
applications  will be the demonstration of ANF for technologically
relevant materials like Gallium or Indium that are among the key
building blocks of modern semiconductor devices.

During ANF the atom-surface interaction and the surface diffusion
play a crucial role in determining the lateral resolution of the
nanostructures. Indeed surface mobility of the adatoms can
increase the structures width or in some cases even wash them out
completely, preventing any direct observation of periodic
nanostructures created through atom focusing. Furthermore, gallium
atoms deposited on substrate are known to self-organize into three
dimensional nano-islands randomly distributed on the surface
substrate.

In this paper we first describe the realization of our gallium
atomic beam and, briefly, the cooling scheme, that is a crucial
step for ANF experiment, then we focus on the analysis of
self-organization properties of gallium atoms deposited on a Si
substrate surface. Our main interest is to understand how the
adatoms self-assembling process could affect the formation of
periodic nano-patterned structures created through atom focusing
technique . Furthermore, the study of  Gallium atoms
self-assembled nanostructures is of great interest in material
research field. In fact, Gallium self assembled nano-islands are
known to be   good precursors for bulk synthesis of Silicon
nanowires~\cite{Sunkara}.

\section{Experimental}
\paragraph{Vacuum system.} The experimental scheme for the vacuum system
(Fig. 1) has been carefully planned according to the following
criteria: a) keep the system as simple and standard as possible
within the current UHV technologies; b) have the possibility to
extend and implement the set-up into a Molecular Beam Epitaxy
(MBE) system; c) allow the optical access needed for the laser
cooling and manipulation of Gallium atoms.\\
The vacuum chamber can be divided in three regions: I) production
of the Gallium atomic beam; II) atom collimation through both
mechanical (skimmer) and optical (laser cooling) means; III)
optical focusing and deposition of the atoms on a substrate. The
chamber is equipped with several viewports AR(Anti-Reflection)
coated for the relevant optical wavelength to allow optical
collimation, focusing and probing of the atomic beam.\\
The atomic source is a  VTS-Createc Gallium effusion cell (Dual
Filament) with a 1~mm insert in its PBN (Pyrolitic-Bore-Nitrate)
crucible. The dual filament configuration prevents the Gallium to
condense, leading to possible damaging of the oven. The effusion
cell is operated at a temperature of about 1000~$^{\circ}$C and
provides an atomic flux of about $5\cdot 10^{14}$~atoms/s
calculated from the Knudsen law. For the deposition of Gallium
nanostructures we need a well collimated high flux atomic beam. To
have a first collimation we geometrically cut the flux using two
skimmers with 1~mm size and 0.8~mm size respectively, placed at a
short distance after the Gallium source. A better collimation
stage imposes the use of laser cooling techniques on the
transverse atomic distribution. The longitudinal mean velocity,
being of the order of $v_L=690~m/s$, is not affected by transverse
laser collimation.\\
The last part of the vacuum system is devoted
to the optical focusing onto a substrate. The cooling and focusing
regions are kept under UHV (better than $10^{-7}$~Pa) conditions
by two ion
pumps.\\
The arrangement described above was set up together with a
complementary optical part (laser sources and optics) to
demonstrate laser cooling and manipulation of gallium atoms for
nanofabrication of ordered structures~\cite{Marago}.
\paragraph{Atomic Physics.}
The Gallium element (Z=31) has two main isotopes $^{69}$Ga
(60.1\%) and $^{71}$Ga (39.9\%), both with nuclear spin I=3/2 and
its electronic configuration leads to a ground state
$4p^2P_{1/2}$. The Gallium atom is a complex atom from the atomic
physicist's point of view and there is no closed transition from
the true ground state suitable for laser cooling. Our choice about
the cooling scheme has been to use the new blue/violet NICHIA
laser diodes at 403~nm and 417~nm to investigate two-colour laser
cooling on the P-S transitions (Fig. 2). Another scheme of laser
cooling at 294~nm (by frequency doubling a Dye laser), based on a
closed transition between the $P_{3/2}$ and $D_{5/2}$ states, is
in use at Colorado State University by the group of Siu Au
Lee~\cite{Rehse}.
\paragraph{Gallium self-assembled nanostructures.}
Depositions of gallium neutral atoms were made on Silicon surfaces
(100), kept at room temperature, without applying the cooling
beams and the standing wave for atom focusing on substrates in
order to test solely their self-assembling properties.\\
The patterned substrates are obtained using standard process
techniques and process integration schemes currently used in
microelectronics~\cite{Chang}. All substrates used for our
experiments have been chemically etched by a solution of HF and
deionized H2O (1:50 at room temperature for 20~s) in order to
remove the native oxidation of the silicon surface.\\
We performed Transmission Electron Microscopy (TEM) analysis of
the deposited samples using a JEOL JEM-2010 field emission
transmission electron microscope operating at 200 KV. The
instrument is also equipped with a GATAN imaging filter and with
an Oxford Instruments LZ5 windowless energy dispersive X-ray
spectrometer (EDS).

\section{Results and discussion}
We investigated experimentally the role of the different
photon-atom interactions provided by the violet 403~nm and blue
417~nm lasers~\cite{Marago} and, within this exploratory work, we
obtained evidence for laser conditioning of gallium atoms, when at
least one laser  acted on the blue side of one of the transitions
shown in fig.~2. These results open up the way for the systematic
study of gallium laser cooling and ANF.

It is beyond the aim of this paper a full description of the
atomic physics results, therefore we focus on the  self- assembled
nanostructures obtained in our deposition experiment from the
unconditioned atom beam.

We needed to understand if the Gallium atoms self assembling
properties could prevent any direct observation of nano-patterning
during an ANF experiment, without applying in the experiment, at
the moment, a standing wave for the atom focusing. For this reason
we have chosen some particular substrate morphologies  made to
enable an effective  variation of the deposition rate on the  the
surface substrate by using silicon substrates with different
trenches. In fact, we varied the substrate surface exposed to the
atomic Gallium incoming flux by modulating the slope of this
surface with respect to the flux direction, as emphasized in the
schemes drawn in  fig.~3 (a and b). In such a way we simulated a
variation of the atomic deposition rate  in a similar fashion to
that expected during an ANF experiment.

In fig.~3 TEM images of Gallium nanostructures self-assembled on
two silicon substrates differently patterned by the trenches are
shown. The aim of the comparison was to characterize the Gallium
deposition on different slopes of the substrate surface, and
therefore to characterize different conditions of deposited
thickness. The depositions were made with an exposition time of 90
minutes keeping the substrates at room temperature . For both
surfaces of Figs.~3(a) and 3(b)  we found that the gallium
nanostructures thickness varies along the substrate  surface as a
function of   where   is the angle of the slope. In spite of
spontaneous organization of 3D islands, this result gives evidence
that in our deposition process the Gallium nanostructures grow in
a limited diffusion regime.  Thus the effective modulation of the
deposition rate along the surface is the dominant factor affecting
the growth. Indeed, in the case of self-assembled nanostructures
an increase of the diffusion barrier for the adatoms from the
centre of an island to its edge occurs~\cite{Nita}.  This
diffusion barrier may be considered the analogous of the
Ehrlich-Schwoebel barrier in homoepitaxy. Therefore we believe
that the self assembly properties of Gallium should not prevent
the direct observation of a periodic modulation at least of the
average dimensions of Gallium nanostructures created through atom
focusing, when a sufficient percentage of the gallium atoms
forming the atomic beam are manipulated by a standing wave.

By TEM analysis it was also possible to characterize the cluster
size distribution at different growth regime (Fig. 4). The Gallium
clusters turn out to grow in the Volmer-Weber
mode~\cite{Sondergard,Zinke}. The nanoparticles are formed after
self-organization of vapour condensing on partially wetting (or
partially drying) substrates. The Gallium melting point is 29.8
$^{\circ}$C, but in the case of sub-micrometric nanostructures
this value decreases hence our deposited gallium consisted of
liquid droplets~\cite{DiCicco}.

The growth condition is represented by the relation:
\begin{center}
\begin{equation}
\Gamma_{sv}-\Gamma_{lv}<\Gamma_{sl}<\Gamma_{sv}+\Gamma_{lv}
\end{equation}
\end{center}
where $\Gamma_{sl},\Gamma_{sv},\Gamma_{lv}$ are the free energies
per unit area of the substrate solid - deposit liquid, substrate
solid-vapour and deposit liquid-vapour interfaces, respectively.
In these cases the growth process follows  different regimes: the
nucleation and growth of individual and immobile droplets, the
static coalescence of droplets when the surface coverage saturates
and the new nucleation of small particles in the area cleaned by
previous coalescence phenomena~\cite{Sondergard}. In fact, by
increasing the deposition time we find a bimodal nano-droplets
size distribution: a bell-shaped part, due to coalescence, and a
tail, caused by renucleation and growth of small particles during
deposition, as shown in fig.~4~(d)-(f) at different deposition
times. By several TEM analysis with energy filtered imaging
technique, as in fig.~5, we also observed that the gallium
nano-islands  were surrounded by native Ga oxide with an average
thickness of about 5~nm . This oxide thickness is determined by
sample exposure to ambient air. The smallest particles formed were
completely oxidized.

\section{Conclusions}
In this paper we have studied the behaviour of self-assembled
Gallium nano-islands on Silicon surfaces (100) in order to assess
possible limitations on the control of nano-structured pattern
created during our future ANF experiment. From the experimental
results we observed that the Gallium self- assembled
nanostructures grow in a limited diffusion regime of the Gallium
atoms on the surface. For this reason the  average dimensions of
the nano-islands followed the flux modulation along the surface.
This experimental evidence points out that the self-assembling
process of the gallium adatoms on the substrate surface will not
prevent the direct observation of  a periodic nano-patterning
created by ANF technique, if  a sufficient  percentage of Gallium
atoms is manipulated by the standing wave. We are now pursuing our
investigation to clarify the minimum percentage of laser
manipulated Gallium atoms in the beam ensuring efficient
nanopatterned depositions.

\section{Acknowledgements}
We wish to thank A. Camposeo, F. Fuso, A. Pimpinelli and S. Trusso
for useful discussions. We are also indebted with M. Vulpio from
STMicroelectronics - Catania CRD for providing us with patterned
Silicon surfaces and for useful discussions.

This work is funded by a FIRB (Fondo per gli Investimenti della
Ricerca di Base) project and by the NANOCOLD project of the IST
Program of the EC through the FET-NID Initiative.



\newpage
\begin{figure}
\center{\scalebox{1}{\includegraphics{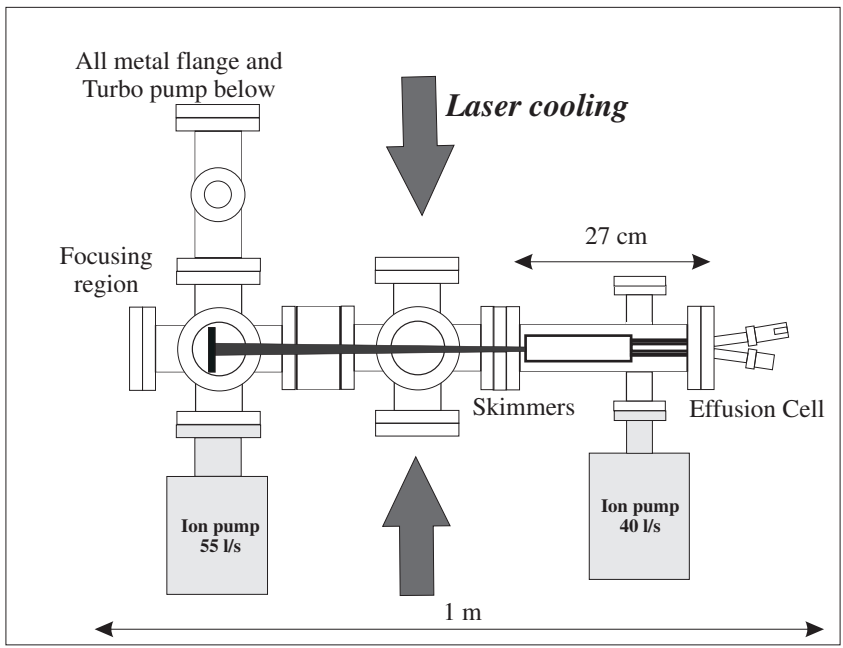}}}
     \caption{Vacuum system for the production of a Gallium atomic beam. }
     \label{vacuum}
\end{figure}

\begin{figure}
\center{\scalebox{0.6}{\includegraphics{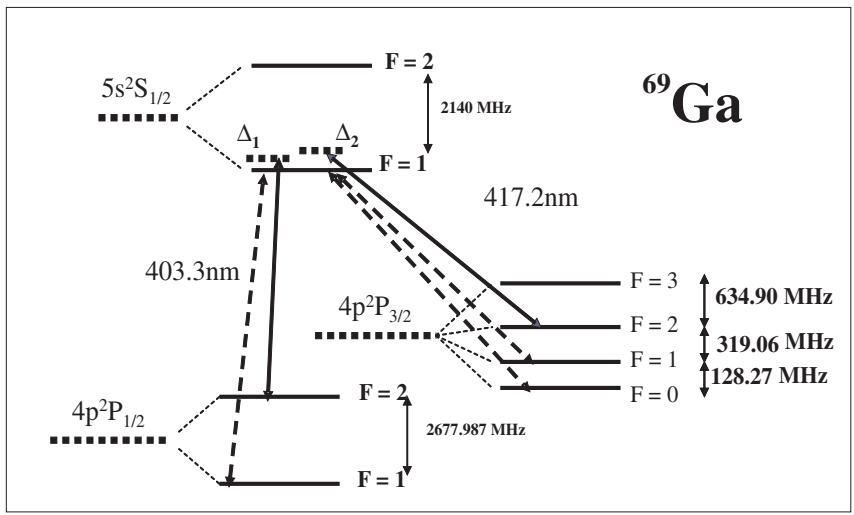}}}
     \caption{Hyperfine splittings for the $^{69}$Ga isotope and a choice of fluorescence cycles for laser
     cooling and repumping.
     To close the cooling transitions from the ground states (P states) we need blue radiation at 403~nm and 417~nm.}
     \label{fig2}
\end{figure}

\begin{figure}
\center{\scalebox{0.6}{\includegraphics{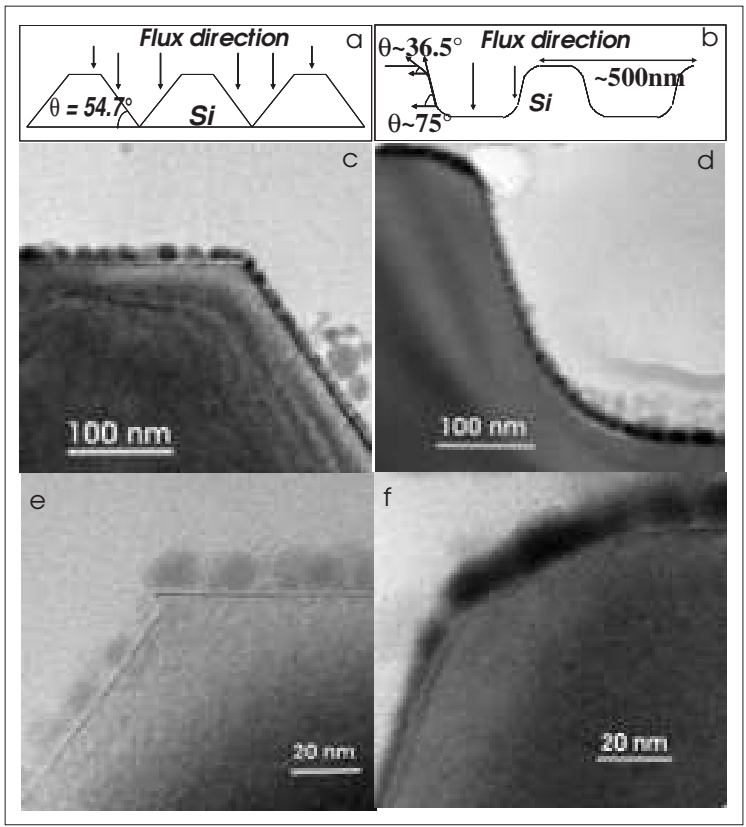}}}
     \caption{In (a) and (b) schemes of the trenches with different slopes realized
     on the silicon substrates are shown. In (c)-(f) are given TEM images in plan view at
     different magnifications of gallium nano-islands self assembled on Si (100) surfaces
     patterned with trenches. The images c), e)  and d), f) correspond  to the two schemes
     for the trenches at different slopes, respectively a) and b).
     In both cases it is evident how the gallium nano-islands thickness changes linearly with
     the slopes following a law.}
     \label{fig3}
\end{figure}

\begin{figure}
\center{\scalebox{0.6}{\includegraphics{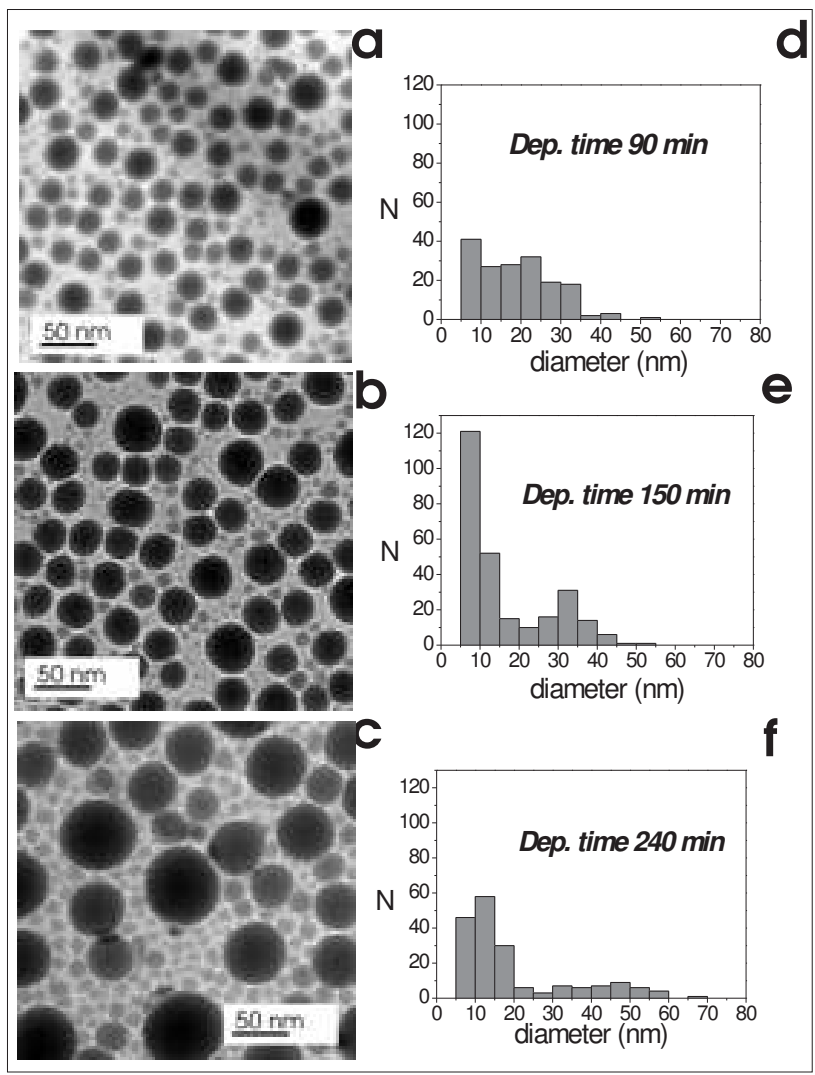}}}
     \caption{In (a), (b) and (c) are shown TEM images in plan view of
     gallium nano-islands self-assembled on Si (100) flat surfaces at
     different deposition times, 90, 150 and 240 minutes respectively. In (d), (e) and (f)
     histograms of the number of nanoparticles N versus the diameter d in nm.}
\label{fig4}
\end{figure}

\begin{figure}
\center{\scalebox{0.6}{\includegraphics{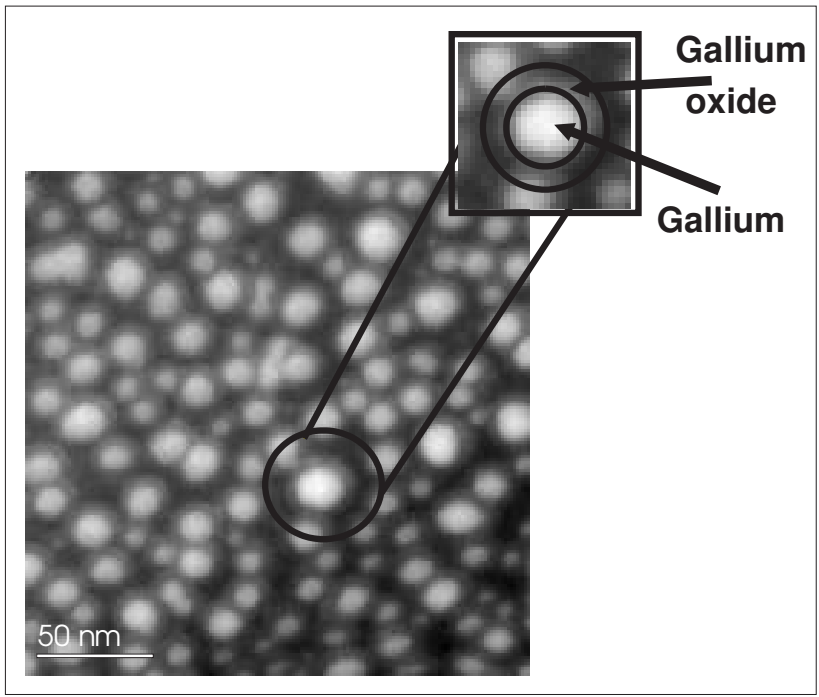}}}
     \caption{SEnergy filtered TEM image of gallium nanoparticles on Si(100)
     formed after 100 min of deposition at room temperature.
     The energy filter (12 eV) was used to evidence the particle structure (core of gallium surrounded by oxide).}
     \label{fig5}
\end{figure}

\end{document}